\documentclass[fleqn,usenatbib]{mnras}

\usepackage[T1]{fontenc}

\DeclareRobustCommand{\VAN}[3]{#2}
\let\VANthebibliography\thebibliography
\def\thebibliography{\DeclareRobustCommand{\VAN}[3]{##3}\VANthebibliography}

\usepackage{graphicx}	
\usepackage{amsmath}	
\usepackage{amssymb}	
\usepackage{hyperref}   

\usepackage{multirow}

\title[Self-supervised kinematic modelling]{A self-supervised, physics-aware, Bayesian neural network architecture for modelling galaxy emission-line kinematics}

\author[James M. Dawson et al.]{
James M. Dawson$^{1}$\thanks{E-mail: dawsonj5@cardiff.ac.uk},
Timothy A. Davis$^{1}$, 
Edward L. Gomez$^{1,2}$,
and 
Justus Schock$^{3}$
\\
$^{1}$Cardiff University, School of Physics and Astronomy, The Parade, Cardiff CF24 3AA, UK\\
$^{2}$Las Cumbres Observatory, Suite 102, 6740 Cortona Dr, Goleta, CA 93117, USA\\
$^{3}$RWTH Aachen University, Templergraben 55, 52062 Aachen, Germany
}

\date{Accepted 2021 February 10. Received 2021 February 9; in original form 2020 October 19}

\pubyear{2021}

\usepackage{newtxtext,newtxmath}

\begin{document}
\label{firstpage}
\pagerange{\pageref{firstpage}--\pageref{lastpage}}
\maketitle

\begin{abstract}
In the upcoming decades large facilities, such as the SKA, will provide resolved observations of the kinematics of millions of galaxies. In order to assist in the timely exploitation of these vast datasets we \textcolor{black}{explore the use of a} self-supervised, physics aware neural network capable of Bayesian kinematic modelling of galaxies. We demonstrate the network's ability to model the kinematics of cold gas in galaxies with an emphasis on recovering physical parameters and accompanying \textcolor{black}{modelling} errors. The model is able to recover rotation curves, inclinations and disc scale lengths for both CO and H\textsc{i} data which match well with those found in the literature. \textcolor{black}{The model is also able to provide modelling errors \textcolor{black}{over learned} parameters thanks to the application of quasi-Bayesian Monte-Carlo dropout}. This work shows the promising use \textcolor{black}{of} machine learning, and in particular self-supervised neural networks, in the context of kinematically modelling galaxies. \textcolor{black}{This work represents the first steps in applying such models for kinematic fitting and we propose that variants of our model would seem especially suitable for enabling emission-line science} from upcoming surveys with e.g. the SKA, allowing fast exploitation of these large datasets.  
\end{abstract}





\section{Introduction}

In studying galaxy evolution, astronomers often use the atomic Hydrogen (H\textsc{i}) 21-cm line to trace the outermost regions of galactic discs (e.g. \citealt{Warren, Begum, Sancisi, Heald, Koribalski2018}). This region can mark the continuous boundary between galaxies and their surrounding environments, including the dark matter halos within which galaxies are thought to reside. The rotation curves of extended H\textsc{i} discs can be used to begin probing the properties of dark matter halos as well as allow the detailed modelling of galaxies' mass distributions when coupled with ancillary observations (e.g. \citealt{Albada, blok}). In the local Universe, H\textsc{i} discs are useful in determining the gaseous content of a galaxy as well as allowing astronomers to probe kinematic properties ranging from substructures such as bars, warps, counter-rotating discs, and spiral arms (e.g. \citealt{jozsa, Spekkens, kamphuis_automated_2015, 3DBAROLO}). Molecular gas observations (typically of the CO molecule) can provide a complimentary view of these regions at high resolution, revealing the interplay between these gas phases. H\textsc{i} is typically more extended than molecular gas, however, allowing it to trace environmental properties such as extended tidal features and the existence of dwarf companions (\citealt{Hibbard, Sancisi, Heald, Serra, bosma_2016, Koribalski2018}).

The evolution of H\textsc{i} gives astronomers insight into the method by which galaxies accrete material from surrounding environments and how the mass of galaxies builds and evolves through star formation. The next generation of H\textsc{i} survey instruments (e.g. the Square Kilometre Array, \citealt{SKA}, Australian Square Kilometre Array Pathfinder, \citealt{Johnston_2007, Johnston_2008}, the South African Meer-Karoo Array Telescope, \citealt{Jonas}, the Chinese Five-hundred metre Aperture Spherical Telescope, \citealt{FAST}) are poised to collect observations spanning a large look-back time, advancing our H\textsc{i} driven science as well as pushing this field of astronomy firmly into the \textit{Big Data} era.

Currently it is estimated that the \textbf{S}quare \textbf{K}ilometre \textbf{A}rray (SKA) will collect data on the order of hundreds of petabytes per year. Given that amount of data is not only too much to fully exploit by hand but also too large to store, astronomers should be looking to develop real-time models that can perform efficient science on incoming data. In an ideal world, physical information would be extracted from incoming data automatically, leaving the work of unravelling the prevailing science to astronomers. However, with such large data volumes and time-intensive techniques how are astronomers to begin moving in a direction in which we can fully exploit the data quality promised by the SKA?

In previous work we sought to begin addressing this challenge via the application of machine learning \citep{dawson_using_2019}, and in particular neural networks, to extract kinematic properties of cold gas in galaxies. Models and tools exist to do this kind of work already. With the upcoming data releases from surveys such as the Widefield ASKAP L-Band Legacy All-Sky Blind Survey (\href{https://wallaby-survey.org/}{WALLABY}), it comes as no surprise that kinematic modelling tools (e.g. \texttt{3D-BAROLO}\footnote{\label{BAROLO}\url{https://editeodoro.github.io/BBarolo/}} \citealt{3DBAROLO}, \texttt{2DBAT}\footnote{\label{2DBAT}\url{https://github.com/seheonoh/2dbat}} \citealt{oh_2d_2018}), \texttt{FAT}\footnote{\label{FAT}\url{https://github.com/PeterKamphuis/FAT}} \citealt{kamphuis_automated_2015}, and \texttt{KinMS}\footnote{\label{KinMS}\url{https://github.com/TimothyADavis/KinMSpy}} \citealt{davis_black-hole_2013, KinMS}) have been in use and ongoing development for some time. Yet these models typically require several minutes or more to provide a full kinematic model of a single object, and longer if errors are required, which may prove problematic for kinematic analyses at SKA survey speeds.

In the past decade machine learning (ML) has become a popular solution to many \textit{Big Data} challenges in galaxy evolution studies (e.g. \citealt{dieleman_rotation-invariant_2015,dominguez_sanchez_improving_2018,dominguez_sanchez_transfer_2018,ackermann_using_2018}, \citealt{Bekki2019}), but remains an under-utilised resource among the galaxy kinematics community. \textcolor{black}{Computer vision, which often utilises ML techniques, has been successfully applied to \textcolor{black}{kinematic characterisation} (e.g. \citealt{Radon}). Yet, there is a distinct absence of directly exploiting ML (with the notable exception of a few recent works, e.g. \citealt{Shen}).} \textcolor{black}{Recently our group has made attempts to exploit the use of ML in this field}, featuring the use of convolutional autoencoders to identify disturbed cold gas in galaxies using data from both simulations and observations (see \citealt{dawson_using_2019}). We still have a long way to go in fully exploring the application of ML to galaxy kinematic characterisation but it appears to be a promising avenue of research and one which we explore further in this work. 

While conventional ML models are capable of high empirical accuracy and low testing time (e.g. \citealt{Random, krizhevsky_imagenet_2012}), they are often highlighted for their slow training times (\citealt{training_time}) and, in some cases, reluctance to generalise to unseen datasets \citep{Dinh,Kawaguchi}. These qualities are unsuitable for survey tasks proposed for the SKA and therefore we are required to look at alternative methods that incorporate the benefits of ML, without the drawbacks associated with standard ML practice.

Such an approach \textcolor{black}{may exist} in the form of self-supervised learning \citep{self_supervised}, whereby models train themselves without the need for an isolated training set. This has huge benefits in that one does not require long training times on a throw-away-dataset, essentially eliminating data wastage. \textcolor{black}{As with all machine learning approaches, self-supervised learning does have its disadvantages including requiring fixed analytical functions to perform training, as well as results which change depending on when one wishes to evaluate test data throughout the model training procedure.} Few pilot tests of these networks exist in astronomy \textcolor{black}{(and even fewer utilising physics-aware capabilities, e.g. \citealt{aragon-calvo_self-supervised_2019})} and none exist in \textcolor{black}{the modelling of galaxy kinematics}. In this paper we present the current results from \textcolor{black}{our first attempts at creating} a self-supervised neural network with the primary goal of inferring the kinematic properties of gas discs in galaxies and an emphasis on \textcolor{black}{extracting (simplistic) characteristics of their rotation curves}. \\

The paper is divided into 3 main sections. \S\ref{sec:Model} gives an in depth description of the model architecture used throughout this work, with emphasis lying on the decoder subnet described in \S\ref{subsec:decoder}. \S\ref{sec:results} presents the results from testing the network using synthetic and real interferometric observations, and \S\ref{sec:conclusion} summarises the main outcomes of the work presented in this paper as well as proposed avenues for future work. 


\section{The model}\label{sec:Model}

\begin{figure}
	\includegraphics[width=\linewidth]{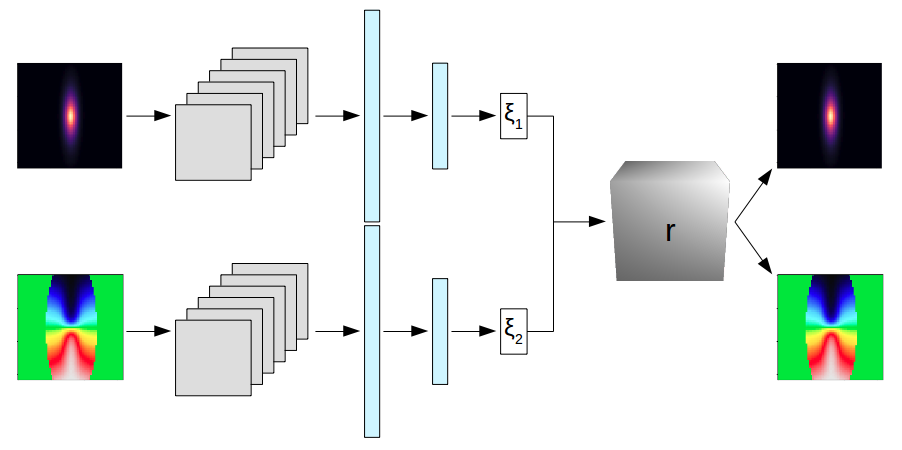}
    \caption{A simplified pictorial representation of the neural network used throughout this work. The model features two convolutional encoder subnets which concatenate learned features before passing them to a decoder subnet. The model receives moment maps as inputs and minimises the loss between decoder-generated moment map outputs and the inputs throughout training. In the diagram grey squares indicate convolutional layers, blue rectangles depict linearly connected layers, and the grey cube represents the auxiliary 3D cube \textcolor{black}{containing the coordinate axes} passed into the network.}
    \label{fig:model}
\end{figure}


\subsection{Input data}\label{sec:input_data}

A typical interferometric observation returns visibilities in a complex plane from which one can obtain a 3D datacube consisting of 2D spatial flux observations separated into discrete channels which correspond to observed frequency. It is this channelisation that allows astronomers to measure the line of sight velocities and hence the kinematic properties of galaxies' gas reservoirs. In practice, one can collapse these datacubes further to create 2D maps that reflect the mean properties of the gas in galaxies. A moment zero (integrated \textcolor{black}{intensity}) map is simply a summation along a cube's frequency/velocity dimension\textcolor{black}{:}

\begin{equation}
\text{Moment zero} = \int\text{I}_{\text{v}} \text{dv} = \sum\text{I}_{\text{v}},
\label{eq:mom0}
\end{equation}

\noindent \textcolor{black}{and a moment one (velocity) map is an \textcolor{black}{intensity} weighted averaging of the line of sight velocities }

\begin{equation}
\text{Moment one} = \frac{\int(\text{v})\text{I}_{\text{v}} \text{dv}}{\int \text{I}_{\text{v}} \text{dv}} = \frac{\sum(\text{v})\text{I}_{\text{v}}}{\sum\text{I}_{\text{v}}}\textcolor{black}{.}
\label{eq:mom1}
\end{equation}

Working directly with the datacubes, or in fact the complex visibilities, would be optimal for any fast pipeline kinematic modelling tool. However, we have chosen to work with moment maps in this work as a first step and to avoid the problems associated with channelised inputs as explained further in \S\ref{sec:conclusion}. It should be noted that, because of our choice to use moment maps, the models described in this work are also suitable to analyse optical IFU maps, \textcolor{black}{as they will be handled similarly by the model described in this work and have been shown to encode kinematic information which can be extracted using both analytical and ML approaches (e.g. \citealt{Radon, Hansen}}). This will be explored further in future work (Dawson et al., in prep).

\textcolor{black}{It should be noted that in this work, we are not making attempts to mitigate the effects of "beam smearing" \citep{Swaters, Blais}. During the recovery of datacubes from complex visibilities, the raw observational datacubes are convolved with a \textit{restoring beam} which effectively encodes the complex visibility plane coverage and is, in some ways, analogous to resolution. It is this convolution step which gives rise to "beam smearing", the effects of which are discussed further in \S\ref{sec:resolution} along with implications for interpreting the model results discussed in this work. Counteracting "beam smearing" will need to be tackled in future work to maximise the effectiveness of models of this type.}


\subsection{Model aim}\label{sec:model_overview}

An autoencoder \citep{autoencoders} is a model composed of two subnets, an \textit{encoder} and a \textit{decoder}.  In an \textit{undercomplete autoencoder} the encoder subnet extracts features and reduces input images to a constrained number of nodes. This so-called \textit{bottleneck} forces the network to embed useful information about the input images into a nonlinear manifold from which the decoder subnet reconstructs the input images and is scored against the input image using a loss function.  

The aim of the model used in this work is to extract \textcolor{black}{semantically meaningful} information from observational data. Typical approaches using a convolutional autoencoder (CAE, \citealt{CAE}) (such as that presented in \citealt{dawson_using_2019}) are powerful for extracting arbitrary (hyperparametric) features that define dataset characteristics. \textcolor{black}{During training, a CAE learns to minimise the difference between input and output tensors rather than the difference between an output and target label (whether this be a continuous or categorical set of target classes). A CAE works similarly to a powerful nonlinear generalisation of principle component analysis (PCA, \citealt{PCA}) whereby it finds a continuous nonlinear latent surface on which input data best lies.} In this work, however, we would like to extract \textcolor{black}{semantically meaningful} parameters of observed systems. In order to achieve this we have combined a convolutional autoencoder with a set of analytical, gradient trackable, functions which \textcolor{black}{approximate the functional forms of observed kinematics of galaxies}. 

The model, \textcolor{black}{known as a \textit{semantic autoencoder} (SAE, \citealt{semantic})}, is a modified CAE created using \texttt{PyTorch}\footnote{\label{pytorch}\url{http://pytorch.org/}} 0.4.1, an open source ML library capable of GPU accelerated tensor computation and automatic differentiation \citep{paszke_automatic_2017}. The model has a neural network architecture suited to self-supervised learning, with additional Bayesian capabilities. Figure \ref{fig:model} shows a simplified pictorial representation of the model architecture.

The encoder subnets extract lower dimensional feature representations from input images \textcolor{black}{(here the integrated \textcolor{black}{intensity} and mean velocity maps as described in \S\ref{sec:input_data})} using a combination of convolutional and linearly connected layers; the decoder then reconstructs the input images from the learned feature representations. In a standard convolutional autoencoder, the decoder would make use of transposed convolution operations, however in this network the decoder is composed of analytical functions written using native \texttt{PyTorch}. \textcolor{black}{This imposes a constraint on the CAE by forcing the network to generate a semantic encoding of the input images. As highlighted by \cite{aragon-calvo_self-supervised_2019}, the decoder function can take any possible form, no matter how representative of the true underlying functions being modelled}. In this way, we can be assured that the encoders are learning semantically meaningful properties of the input images and are no longer tied to traditional training methods, instead allowing the network to train on all available data (including test data) in a self-supervised manner. \textcolor{black}{An SAE becomes physics-aware once the assumption is made that the decoder function can be used to reveal physically meaningful information about the input. In this paper, the physics-awareness of the model refers to our main focus of approximating parameterisations for rotation curves, intensity profiles and recovering galaxy inclinations (see \S\ref{subsec:decoder}).}

\textcolor{black}{For a more in-depth background to the use of autoencoders we refer the reader to \cite{Bourlard} and \cite{Hinton504}. For both a concise and thorough introduction to the use of self-supervised, physics aware, neural networks in astronomy we recommend \cite{aragon-calvo_self-supervised_2019}.}

\begin{table}
\centering
\caption{The SAE encoder subnet architecture used throughout this paper. The first column lists the name of each layer/operation, the second column describes the type of layer/operation, the third column shows the dimensions of each layer's output tensors (hence the input shape to the next layer). The dimensions follow the PyTorch convention (batch size, number of channels, height, width). The filter column shows the dimensions (height, width) of kernels used to perform the convolution and pooling operations. \textcolor{black}{The convolutional and linearly connected layer groups are separated by a blank row for clarity.}}
\begin{tabular}{llcc}
\hline
\hline
Name   & Layer/Operation      & Dimensions    & Filter \\ \hline
       &                      &               &        \\
Input  & --                   & (64,1,64,64)  & --   \\
Conv   & 2D Convolution       & (64,16,64,64) & (3,3)  \\
Pool   & 2D Max Pooling       & (64,16,32,32) & (2,2)  \\
Conv   & 2D Convolution       & (64,32,32,32) & (3,3)  \\
ReLU   & ReLU                 & --            & --     \\
Pool   & 2D Max Pooling       & (64,32,16,16) & (2,2)  \\
Conv   & 2D Convolution       & (64,64,16,16) & (3,3)  \\
ReLU   & ReLU                 & --            & --     \\
Pool   & 2D Max Pooling       & (64,64,8,8)   & (2,2)  \\
Conv   & 2D Convolution       & (64,128,8,8)  & (3,3)  \\
ReLU   & ReLU                 & --            & --     \\
Pool   & 2D Max Pooling       & (64,128,4,4)  & (2,2)  \\
       &                      &               &        \\
Lc1    & Linear               & (64,1,1,2048)      & --     \\
ReLU   & ReLU                 & --            & --     \\
Drop   & Dropout (p=0.1)      & --            & --     \\
Lc2    & Linear               & (64,1,1,256)  & --     \\
Htanh  & Hard tanh activation & --            & --     \\
Output & --                   & (64,1,1,2)   & --     \\ \hline
\end{tabular}
\label{table:architecture}
\end{table}


\subsection{The encoder subnets}\label{subsec:encoder}

Within the network, the encoders are two convolutional-classifier-like subnets. Each comprises a series of 4 convolutional and 2 fully connected layers, interspersed with pooling layers and activation functions. The encoders are used to extract and dimensionally reduce features from input images. The two subnets \textcolor{black}{independently} receive a moment zero map (a 2D \textcolor{black}{intensity} profile, \textcolor{black}{normalized} in the range 0--1) and a moment one map (a 2D velocity profile, \textcolor{black}{normalized} into the range -1--1) respectively. \textcolor{black}{Throughout this work, we ensure that the input maps have size of 64$\times$64 pixels. All input maps whose sizes are larger or smaller, like those discussed in \S\ref{sec:VLA} and \S\ref{sec:WISDOM}, are subsequently up/down-sampled to a size of 64$\times$64 using \texttt{PyTorch}'s \texttt{torch.nn.Upsample} class, in \textit{bilinear} mode.} Each moment map carries valuable information for the decoder functions as described in \S\ref{subsec:decoder}. With this in mind, the output of the encoders are two vectors which are concatenated before passing to the decoder subnet. For an in depth look at the encoder subnet structure see Table \ref{table:architecture}. 

The encoders learn the following properties: \textit{subnet 1}: observed galaxy inclination (i) \textcolor{black}{and free parameters of the \textcolor{black}{intensity} profile which make up $\xi_{1}$ in Figure \ref{fig:model}; \textit{subnet 2}: the parameters of the velocity profile of the galaxy which make up $\xi_{2}$ in Figure \ref{fig:model}}.


\subsection{The decoder subnet}\label{subsec:decoder}

Here we detail the functions required for reconstructing the moment zero and moment one input maps from the concatenated feature representations $\xi_{1}$ and $\xi_{2}$ as shown in Figure \ref{fig:model}. 
In recovering the moment maps, we are primarily interested in modelling two profiles. Firstly, the \textcolor{black}{intensity}\textcolor{black}{:}

\begin{equation}\label{eq:SBProf}
\text{I(r)} = \text{I}_0 \, \text{exp}\left({-\frac{\text{r}_{\text{x,y}}}{\text{r}_{\text{scale}}}}\right) \, \text{exp}\left({-\frac{z}{\text{r}_{\text{z-scale}}}}\right) ,
\end{equation}

\noindent where I$_{\text{0}}$ is the intensity normalisation factor (set to 1 throughout, due to the global normalisation described above), r$_\text{x,y}$ is the radius in the $xy$ plane, in arcseconds, r$_{\text{scale}}$ is the \textcolor{black}{intensity} scale length in the $xy$ plane, $z$ is the position in the $z$ axis, and r$_{\text{z-scale}}$ is the \textcolor{black}{intensity} scale length in the $z$ axis set to a value of 1 spaxel throughout this work, to emulate a thin disk. Intensity values are determined by combining the integrals of Equation \ref{eq:SBProf} across each spaxel in the $xy$ and $z$ planes.

Secondly, the rotational velocity\textcolor{black}{:}

\begin{equation}\label{eq:LOSVel}
\text{V(r)} = \frac{2\text{V}_\text{max}}{\pi} \, \text{arctan} \left( -\frac{r}{\text{r}_{\text{turn}}}\right) ,
\end{equation}

\noindent where V$_{\text{max}}$ is the asymptotic line of sight velocity, r is the radius in arcseconds, and r$_{\text{turn}}$ is the velocity profile scale length.    

\textcolor{black}{Here, our choice of exponential \textcolor{black}{intensity} profile and arctan velocity profile are entirely arbitrary (i.e. not driven by any physical theory), but are choices \textcolor{black}{motivated by some of the simplest forms that can approximately fit the typical disks and rotation curves found in the Universe}. Clearly objects that do not follow these functional forms will not be appropriately fit by this network and we discuss this further in \S\ref{sec:caveats}. \textcolor{black}{However, it should be noted that this analytical-style decoder implementation} would be equally valid for other functional forms. For example, one could choose to fit bulge-disk models with such an architecture, or include the influence of central point masses or the effects of dark matter halos. These more realistic networks will be explored in future works.}

An auxiliary 3D tensor of radii (labelled \textit{r} in Figure \ref{fig:model}) is passed into the network, cloned, and evaluated using Equations \ref{eq:SBProf} and \ref{eq:LOSVel}. The 2D moment maps are then created using Equations \ref{eq:mom0} and \ref{eq:mom1}. \textcolor{black}{The velocity profile is later converted into line of sight velocity map via an inclination projection and velocity weighting based on the pixel angles about the line of sight axis.}


\begin{figure*}
\includegraphics[width=\linewidth]{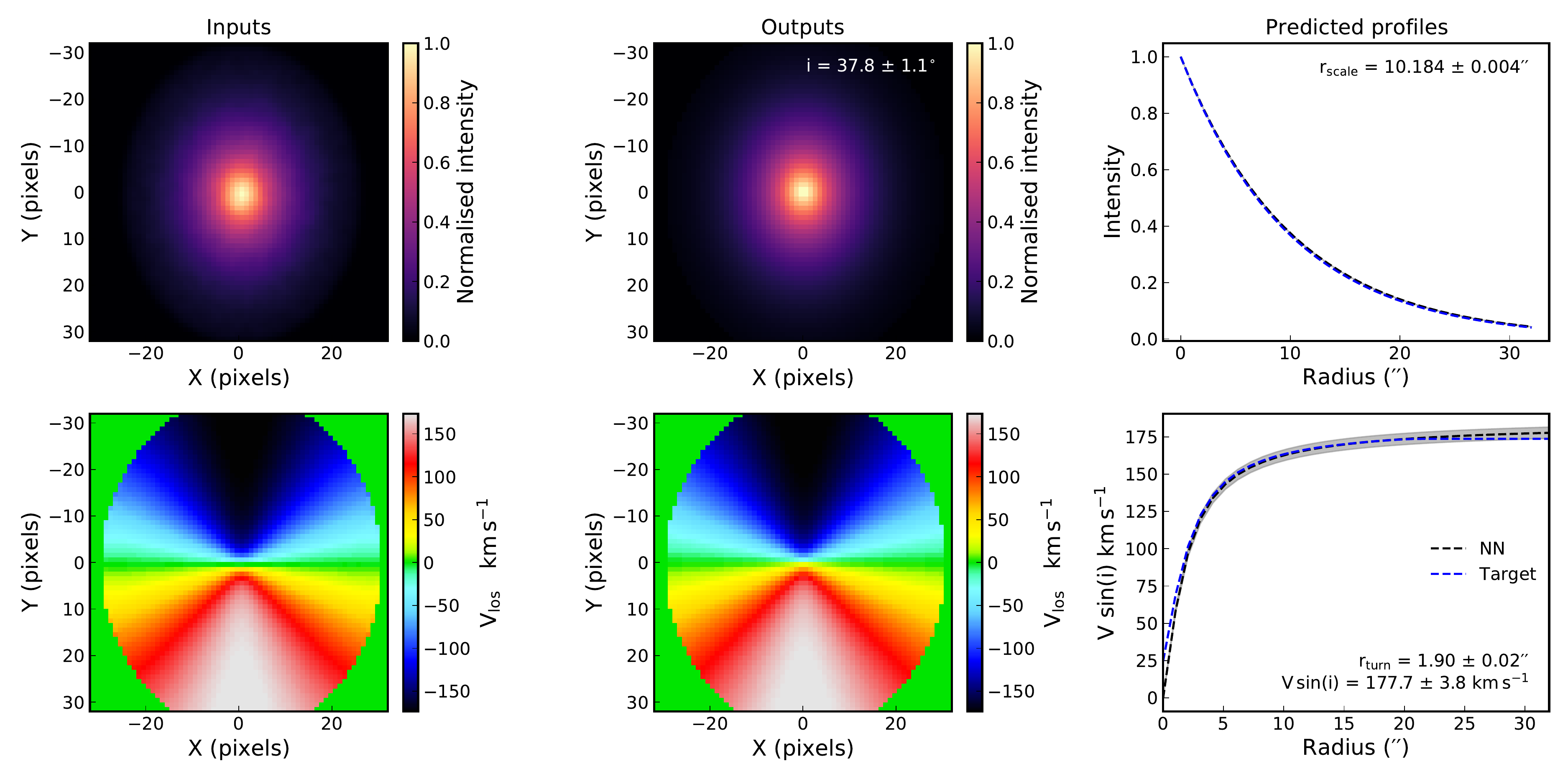}
\caption{A randomly selected synthesised galaxy, created using \texttt{KinMS} and \textcolor{black}{evaluated using the network in \textit{blind} testing mode}. The black dashed lines and grey areas show the mean and 1$\sigma$ modelling uncertainties respectively for profiles predicted by the neural network model. The blue dashed lines show the target profiles which were used to create the input maps. The galaxy was created with the following known parameters: i$=37.2^{\circ}$, r$_\text{scale} = 10.0^{\prime\prime}$, r$_\text{turn} = 1.6^{\prime\prime}$, and V$_{\text{max}}\,$sin(i) $= 173.6$ km$\,\text{s}^{-1}$. \textcolor{black}{The network predicted parameters are shown as text in the upper-middle, upper-right, and lower-right subplots.}}
\label{fig:synthetic_results}
\end{figure*}

\subsection{Model training procedure}\label{sec:training}

The network is trained with minimal optimisation of hyperparameters in order to demonstrate the simple nature of this architecture. At all times the network utilises a \texttt{PyTorch}'s \texttt{MSELoss} function which computes the mean squared error\textcolor{black}{:}

\begin{equation}\label{eq:MSELoss}
\mathcal{L} = \frac{1}{\text{N}} \sum\limits_{\text{i}=0}^{\text{N}} \left(f\left(\text{x}_{\text{i}}\right)-\text{y}_{\text{i}}\right)^{2}, 
\end{equation}

\noindent between the model outputs, y$_{\text{i}}$, and inputs, x$_{\text{i}}$, for every forward pass of a batch of size N. In this case, this is the squared difference between the moment zero and moment one inputs and decoder generated outputs. It is worth noting here that all synthetically generated moment maps have the same position angle and consequently any observational data used for training and testing have been de-rotated using published position angle measurements. We do this as position angle is a non-physical parameter which we can easily account for in preprocessing \textcolor{black}{(with e.g. the \texttt{fit\_kinematic\_pa} routine of \citealt{krajnovic})}. 

We use an adaptive Adam learning rate optimiser \citep{kingma_adam:_2014} which begins with a value of $10^{-4}$ and reduces via multiplication of 0.975 every 2 epochs. We find that the model converges well after 300 epochs for all training runs presented in this paper. 

Where synthetic training data is used, the network receives batches of 64 input moment map pairs. Initial tests showed the network to be largely unaffected by batch size and so 64 is arbitrarily chosen to increase training speed. 

The models and Python training scripts used for the work presented in this paper are publicly available on GitHub\footnote{\label{GitHub}\url{https://github.com/SpaceMeerkat/Corellia/}}.


\subsection{Model testing procedure}\label{sec:testing}

Testing the network can be done in three distinct ways, depending on the situation at hand. In order to test data, one can choose whether to train the network on the test data alone (we call this testing procedure \textit{solo} testing), to train on the test data alongside other examples (we call this testing procedure \textit{combined}), or to use the network in full test mode having only trained on examples not including those data that we wish to test (called \textit{blind} testing).

One can imagine the case where sufficient training data has been passed through the network in a survey, such that in order to return rapid kinematic modelling of new observations one simply passes the new observations through the network with no prior exposure to the training procedure. This \textit{blind} testing has the advantage of rapid testing speed but at the potential cost of lowered predictive accuracy, \textcolor{black}{in an epistemic uncertainty dominated regime}. One can also imagine the case whereby initial survey data has been collected and some sample of the dataset the network used to train is also in need of testing. As the network has seen \textcolor{black}{these} data during the training procedure, \textit{combined} testing has the advantage of potentially higher accuracy at the expense of time needed to train the model. It should come as no surprise that the ideal testing scenario for this network is \textit{combined}, with a sufficiently large training set in an \textcolor{black}{aleatoric uncertainty dominated regime}. However, there are cases (such as at first light of a survey) where the only test data available is that which the network was trained on. It is in this scenario that \textit{solo} testing will occur and although this testing regime lacks the benefits afforded by \textcolor{black}{\textit{combined}} testing, it has the potential advantage of predictions not being influenced by anomalous data whose population increases with training set size.


\subsection{\textit{Monte Carlo} dropout}

In this section we summarise the use of \textit{Monte Carlo} dropout (henceforth MC dropout; \citealt{NIPS}) to provide quasi-statistical modelling uncertainties over learned parameters within the model.

In conventional neural network training circumstances \textit{dropout} may be interpreted as permuting a trained model \citep{JMLR} via the probabilistic zeroing of weights in linearly connected layers. \textcolor{black}{Traditionally, dropout layers are used throughout training} in order to force the network to behave as an ensemble of architectures increased testing accuracy and generalisation power. In the case of MC dropout, after training, dropout is reapplied to the network in evaluation mode and inputs are passed through the model many times, effectively sampling a posterior where the model architecture is marginalised out. \textcolor{black}{\cite{Gal2016Uncertainty} first proposed the idea of approximating distributions over parameters learned in neural networks in this way and has since been used in astronomy (e.g. for the probabilistic labelling of galaxy morphologies, \citealt{galaxy_zoo}).}

For an input x (comprised of a moment 0 and moment one map), training data $\mathcal{D}$, model weights w, T forward-pass evaluations, and encoder output k, the predicted parameter means and standard deviations are given by Equations \ref{eq:params_mean} and \ref{eq:params_std} respectively.

\begin{equation}
\hat{\text{k}} = \frac{1}{\text{T}} \sum_{\text{t}} P(\text{k}|\text{x},\text{w}_\text{t}) 
\label{eq:params_mean}
\end{equation}

\begin{equation}
\sigma = \frac{1}{\text{T}} \sum_{\text{t}} |\text{k}-\text{k}_\text{t}|
\label{eq:params_std}
\end{equation}

For a comprehensive derivation of Equations \ref{eq:params_mean} and \ref{eq:params_std}, as well as the implications for using an arbitrary dropout probability, \textcolor{black}{we refer the reader to \cite{galaxy_zoo}.} Examples of the posterior distributions, p(k|w,$\mathcal{D}$), over learned parameters using MC dropout for a randomly selected synthesised galaxy are described further in \S\ref{subsec:Input_Output}. 

It should be noted that, as the network does not use dropout to zero weights in the convolutional layers, $\sigma$ does not represent a complete error over learned parameters. Instead one should consider $\sigma$ as a lower limit error over parameters whose use becomes immediately obvious for pipeline flagging purposes or to generate relative errors within a test set. \textcolor{black}{The errors produced through this technique are strictly errors due to the modelling technique, and will underestimate the true error in any parameter, which arises due to both modelling and observational uncertainties. }


\section{Results and discussion}\label{sec:results}

In this section we present exemplar test results for highly spatially resolved galaxy observations. In each case we have trained new networks using the procedures described in \S\ref{sec:training}.    


\subsection{Synthesised examples}\label{sec:synthesised}


\subsubsection{Input-output}\label{subsec:Input_Output}

\begin{figure*}
\includegraphics[width=0.8\linewidth]{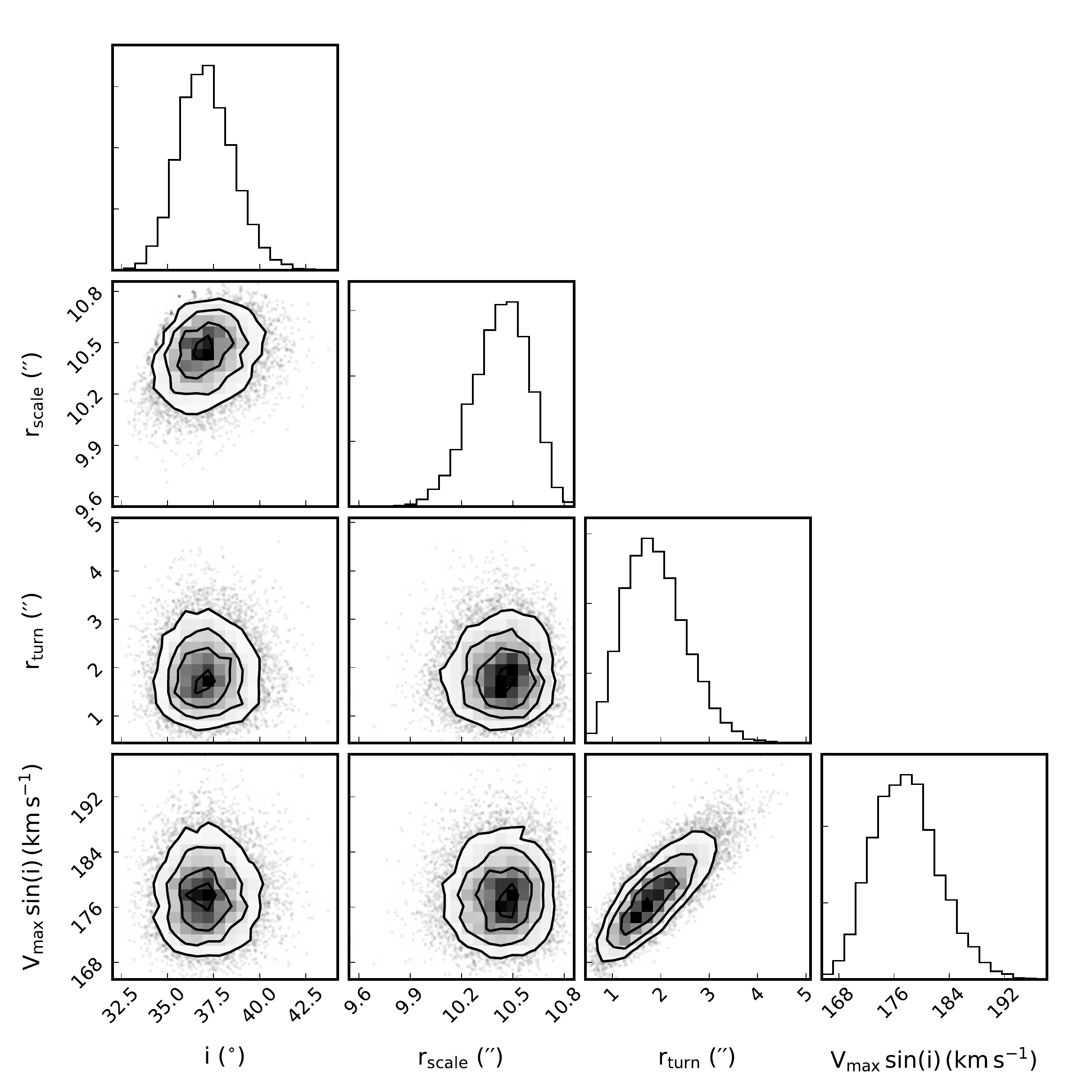}
\caption{\textcolor{black}{Corner plot showing the level of covariance between learned parameters for one randomly generated, synthesised galaxy (discussed \textcolor{black}{further} in \S\ref{sec:synthesised}). The accompanying histograms represent quasi-probabilistic distributions thanks to the use of \textit{Monte Carlo} dropout. This galaxy was passed through the network in test mode $10\,000$ times in order to build the distributions. We observe well constrained learned parameters with Gaussian like profiles, allowing for quasi-probabilistic modelling errors for the parameters. The only strong covariance observed is that between the maximum line of sight velocity and the velocity profile scale length, which is entirely expected and present in traditional kinematic analyses.}}
\label{fig:MC_Dropout}
\end{figure*}

\begin{figure*}
\includegraphics[width=0.7\linewidth]{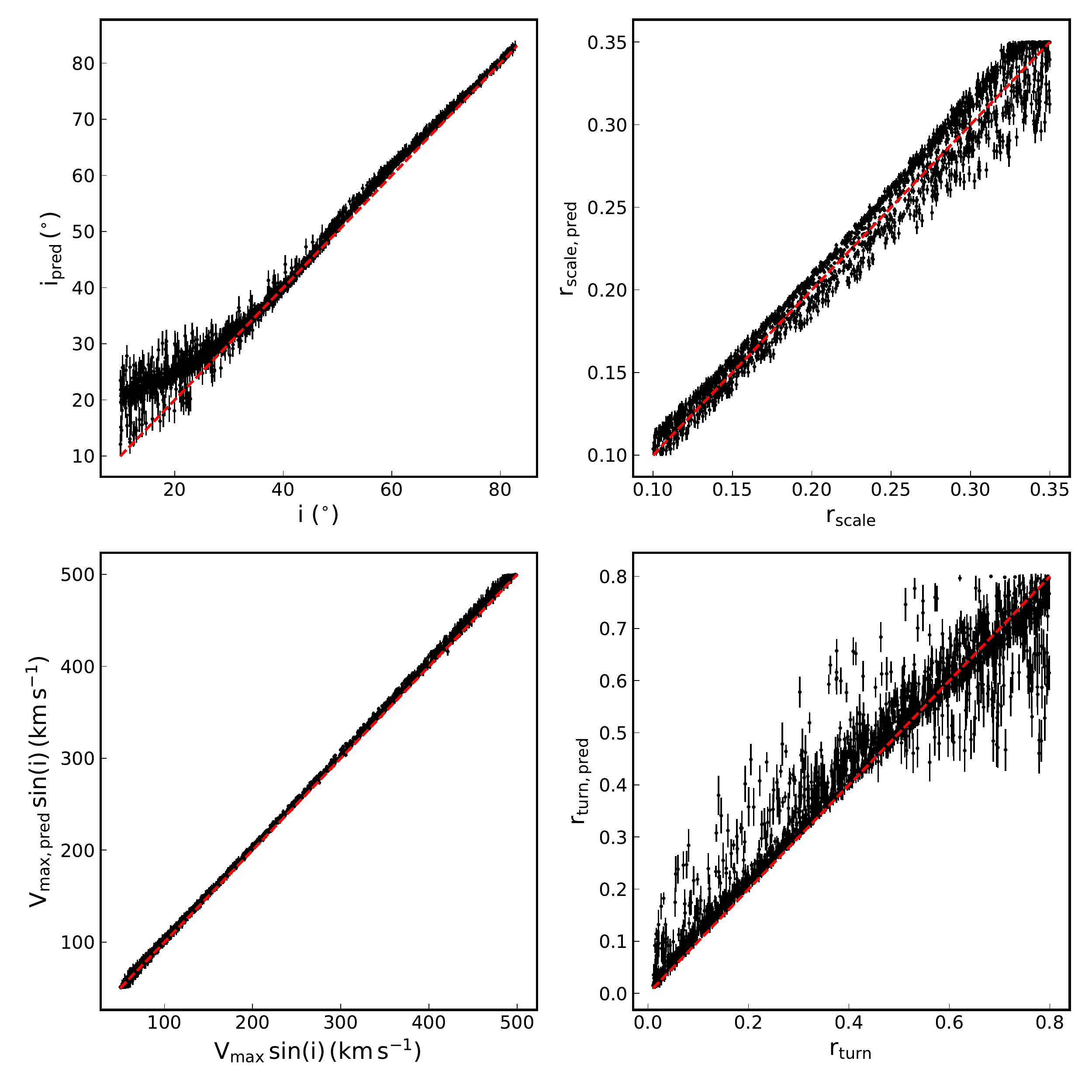}
\caption{True versus predicted plots for each learnable parameter in the network. Black markers and error bars pertain to the tested galaxies and the red dashed line indicates the 1:1 line on which perfect predictions should ideally lie. This model was trained using purely synthetic data with a restoring beamsize of \textcolor{black}{2 resolution elements and only including well resolved examples as discussed in \S\ref{sec:synthesised}}. Those galaxies whose projected r$_\text{turn}$ fell below 1.5 times the restoring beamsize were removed in order to mimic the automated flagging of poorly resolved galaxies at high inclination in a survey. Of the 2000 synthesised galaxies tested, 261 (13$\%$) were removed using this cut.}
\label{fig:synthetic_cut}
\end{figure*}

\begin{table}
\centering
\caption{Parameter values and ranges for all synthetically generated galaxies using the \texttt{KinMS} package. \textcolor{black}{The units for r$_{\text{scale}}$ and r$_{\text{turn}}$ are absent due to both quantities being fractions of the input map size.} The position angle of each galaxy is fixed at 0 as it is not a physically meaningful parameter. Throughout model training, parameters are drawn uniformly in the ranges listed.}
\begin{tabular}{lcl}
\hline
\hline
Parameter               & Size/range  & Units \\ \hline
                       &             &       \\
Position angle         & 0           & deg   \\
Inclination            & 10--90    & deg   \\
r$_\text{scale}$           & 0.1--0.35 & --    \\
r$_\text{turn}$           & 0.01--0.8 & --    \\
V$_\text{max}\,$sin(i) & 50--500   & km$\,\text{s}^{-1}$  \\ \hline
\end{tabular}
\label{table:KinMS}
\end{table}

In order to explore the limitations of the network, we tested the model using synthetic galaxies generated using the \texttt{Python} based kinematic simulator \texttt{KinMS}\footnote{\label{kinms}\url{https://github.com/TimothyADavis/KinMSpy}} (KINematic Molecular Simulation, \citealt{davis_black-hole_2013, KinMS}). Figure \ref{fig:synthetic_results} shows the inputs and outputs as well as both known and predicted profiles for a \textcolor{black}{galaxy generated using the same analytical functions described in \S\ref{subsec:decoder} with inclination, maximum velocity, and scale lengths drawn randomly in the ranges shown in Table \ref{table:KinMS}}, \textcolor{black}{and a fixed beam size of 2 resolution elements}. It is clear that the model is able to recover the galaxy's rotation curve (and other parameters) well in \textit{blind} testing mode, \textcolor{black}{whereby the model has not yet trained on the test data}. \textcolor{black}{The quasi-probabilistic distributions for each learned parameter for this galaxy are shown in Figure \ref{fig:MC_Dropout}, highlighting the Gaussian-like nature of the learned parameter distributions as well as an expected covariance between r$_{\text{turn}}$ and V$_{\text{max}}\,$sin(i).}

As seen in Figure \ref{fig:synthetic_cut} the model is able to recover the desired physical parameters of synthesised galaxies well, heuristically. \textcolor{black}{For the} 1739 test galaxies shown in Figure \ref{fig:synthetic_cut} we measure the average deviation of parameters: i, r$_{\text{scale}}$, r$_{\text{turn}}$, and V$_{\text{max}}$sin(i), from the 1:1 line as $\sigma_{\text{i}} = 0.98^{\circ}$, $\sigma_{\text{r}_{\text{scale}}}$ = 0.003, $\sigma_{\text{V}_{\text{max}}\text{sin(i)}} = 3.48\,\text{km}\,\text{s}^{-1}$, and $\sigma_{\text{r}_{\text{turn}}}$ = 0.017 respectively. 

\textcolor{black}{It is clear from Figure \ref{fig:synthetic_cut} that the error estimates do not represent the total errors over the parameters and only encode the modelling error. This makes the presented errors strictly lower limit estimates, and mostly useful for comparing reliability within the dataset, rather than external use. This can be seen by the fact that on average only $\sim$35\% of the data points in Figure \ref{fig:synthetic_cut} have errorbars which overlap with the 1:1 true-versus-predicted line. For the presented dataset these errors likely underestimate the total error by a factor of $\sim$2.5. Including errors in the observations themselves will help to narrow this gap and will be explored further in future work}.


\subsubsection{The effect of resolution}\label{sec:resolution}

One expects r$_\text{scale,pred}$ to artificially increase with beam size for a fixed r$_{\text{scale}}$. However, r$_{\text{scale}}$ is not known for observations of galaxies whose values r$_{\text{scale}}$ fall below some fraction of the beamsize. We see this effect happening as shown in Figure \ref{fig:beam_offsets} in a non-complex manner. Therefore, we recommend enforcing flagging based on inclination which appears to be strongly linked with those galaxies whose r$_{\text{scale}}$ is under predicted (along the minor axis). In the edge-on galaxy case, the minor axis is no longer well resolved resulting in a poor recovery of the \textcolor{black}{intensity} profile. \textcolor{black}{However, this is a well-known issue in moment based kinematic modelling, in which the intensity profiles and kinematics can never be fully derived in edge-on galaxies due to line of sight effects.}

As we have included no method for mitigating the changes induced by varying beam size, it comes as no surprise that the network will behave differently given a sufficiently large ratio of beam size to galaxy extent. \textcolor{black}{Given that we do not have a mechanism for dealing with "beam smearing" in the current network architecture, we expect to see its influence, lowering the apparent line of sight velocities close to the center of galaxies where the \textcolor{black}{iso-velocity contours} are closest together. For minimising the effects of varying beam size \textcolor{black}{we recommend convolving the 3D spatial cube $r$ (see Figure \ref{fig:model}), evaluated using Equation \ref{eq:SBProf}, with the restoring beam before creating the output maps}. The advantage of this approach being that the restoring beam is often included in data-product header units, and so should be readily available for creating kernels with which to perform the aforementioned convolution. We consider this approach as beyond the scope of the work presented in this paper, but will be included in future work focusing on retrieving the properties of marginally resolved galaxies.}

\subsubsection{Fill factor}

In previous work we showed that the fill factor (i.e. the number of zeroed pixels) in a velocity map's field of view, impacts the behaviour of NN models which take them as inputs \citep{dawson_using_2019}. With the NN model presented in this work, we have seen little evidence that this has an effect on the galaxies' predicted \textcolor{black}{parameters}. We attribute this behaviour to the nature of the training procedure, whereby in \textit{combined} and \textit{solo} testing, the network does not rely solely upon inference of unseen data. 


\begin{figure*}
\includegraphics[width=\linewidth]{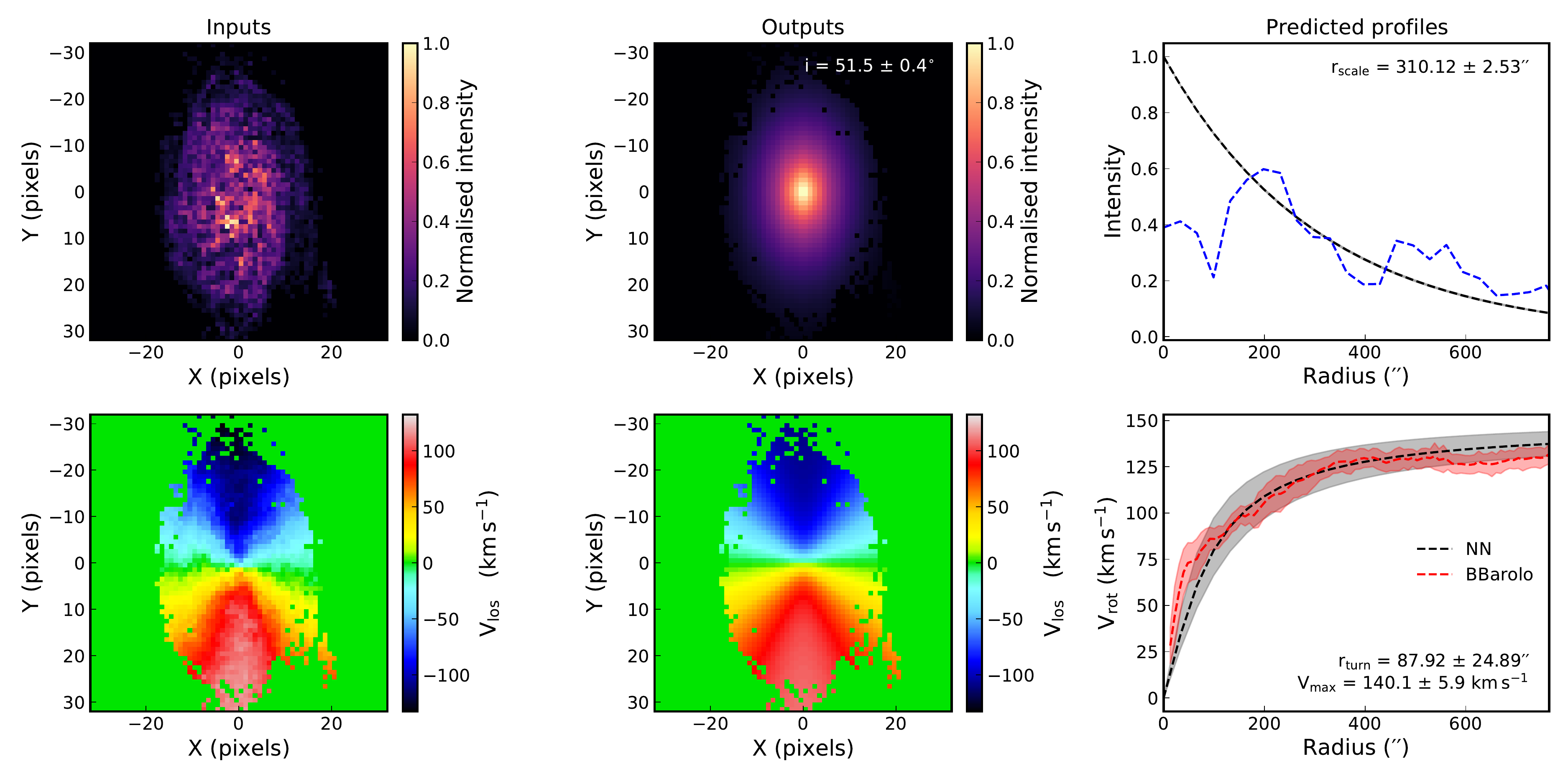}
\caption{\textcolor{black}{An example galaxy, NGC 2403, observed in H\textsc{i} and evaluated using the network in \textit{combined} testing mode. Maps in the left and middle columns share $x$ and $y$ axis sizes of 64$\times$64 pixels. In this way we are directly observing the input and output maps of the model. The right column has undergone an $x$-axis rescaling to match observational scales found in the literature. The black dashed lines and grey areas show the mean and 1$\sigma$ modelling errors respectively for profiles predicted by the neural network. The blue dashed line shows a major axis cut of the input intensity map. The red dashed line and filled area show the best fit and associated errors modelled using \texttt{BBarolo} on the datacube. In order to make a direct comparison between the network's and \texttt{BBarolo}'s derived rotation curves, the network's velocity profile has been corrected for by the predicted inclination term. \textcolor{black}{The network predicted parameters are shown as text in the upper-middle, upper-right, and lower-right subplots.} We see that this galaxy has a velocity profile which can be roughly approximated by an arctan function meaning the kinematic parameters are well recovered by the model.}}
\label{fig:NGC2403}
\end{figure*}

\begin{figure*}
\includegraphics[width=\linewidth]{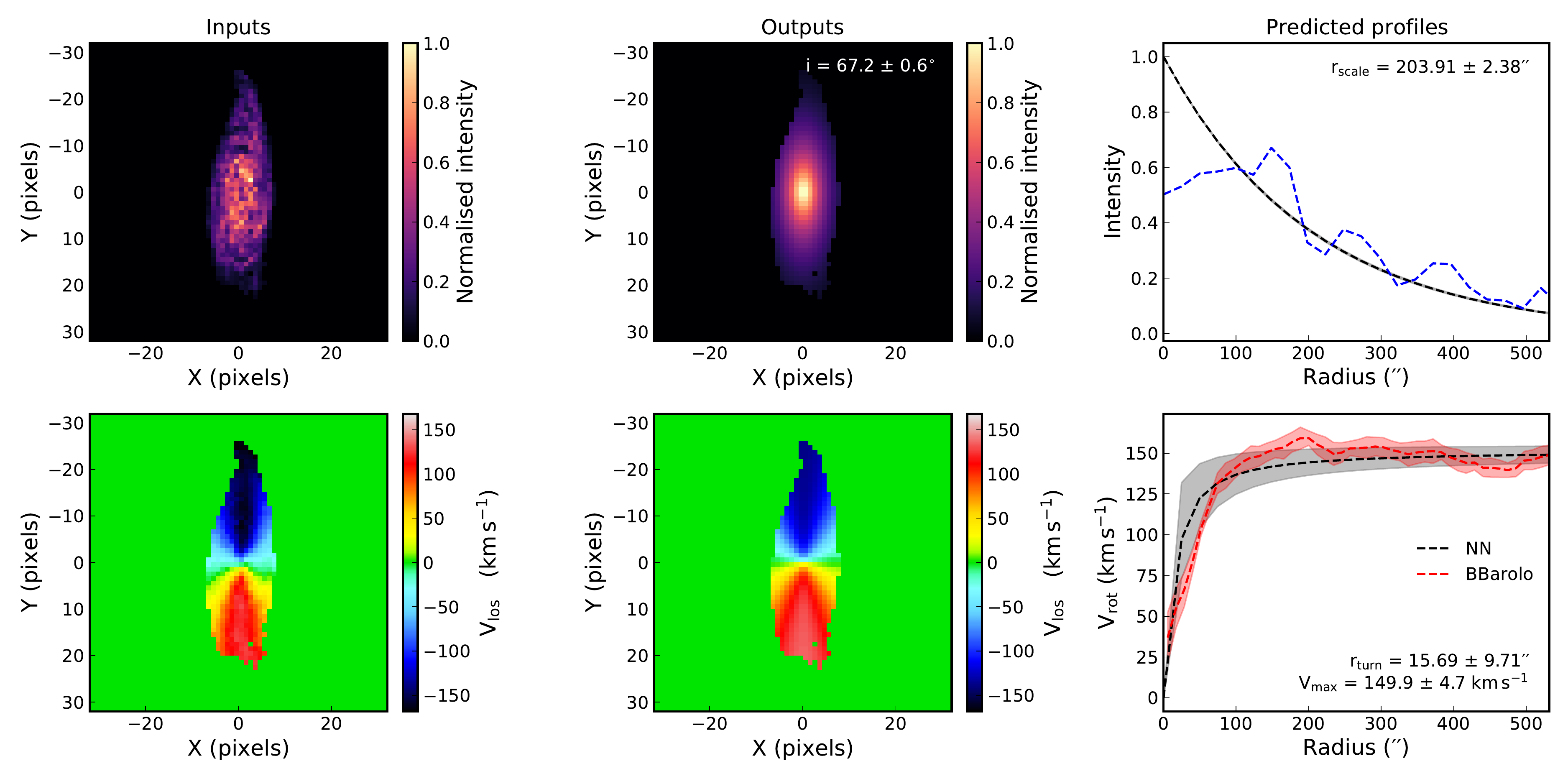}
\caption{\textcolor{black}{An example galaxy, NGC 3198, observed in H\textsc{i} and evaluated using the network in \textit{combined} testing mode. Maps in the left and middle columns share $x$ and $y$ axis sizes of 64$\times$64 pixels. In this way we are directly observing the input and output maps of the model. The right column has undergone an $x$-axis rescaling to match observational scales found in the literature. The black dashed lines and grey areas show the mean and 1$\sigma$ modelling errors respectively for profiles predicted by the neural network. The blue dashed line shows a major axis cut of the input intensity map. The red dashed line and filled area show the best fit and associated errors modelled using \texttt{BBarolo} on the datacube. In order to make a direct comparison between the network's and \texttt{BBarolo}'s derived rotation curves, the network's velocity profile has been corrected for by the predicted inclination term. \textcolor{black}{The network predicted parameters are shown as text in the upper-middle, upper-right, and lower-right subplots.} We see that this galaxy has a velocity profile which can be roughly approximated by an arctan function meaning the kinematic parameters are well recovered by the model.}}
\label{fig:NGC3198}
\end{figure*}

\begin{figure*}
\includegraphics[width=\linewidth]{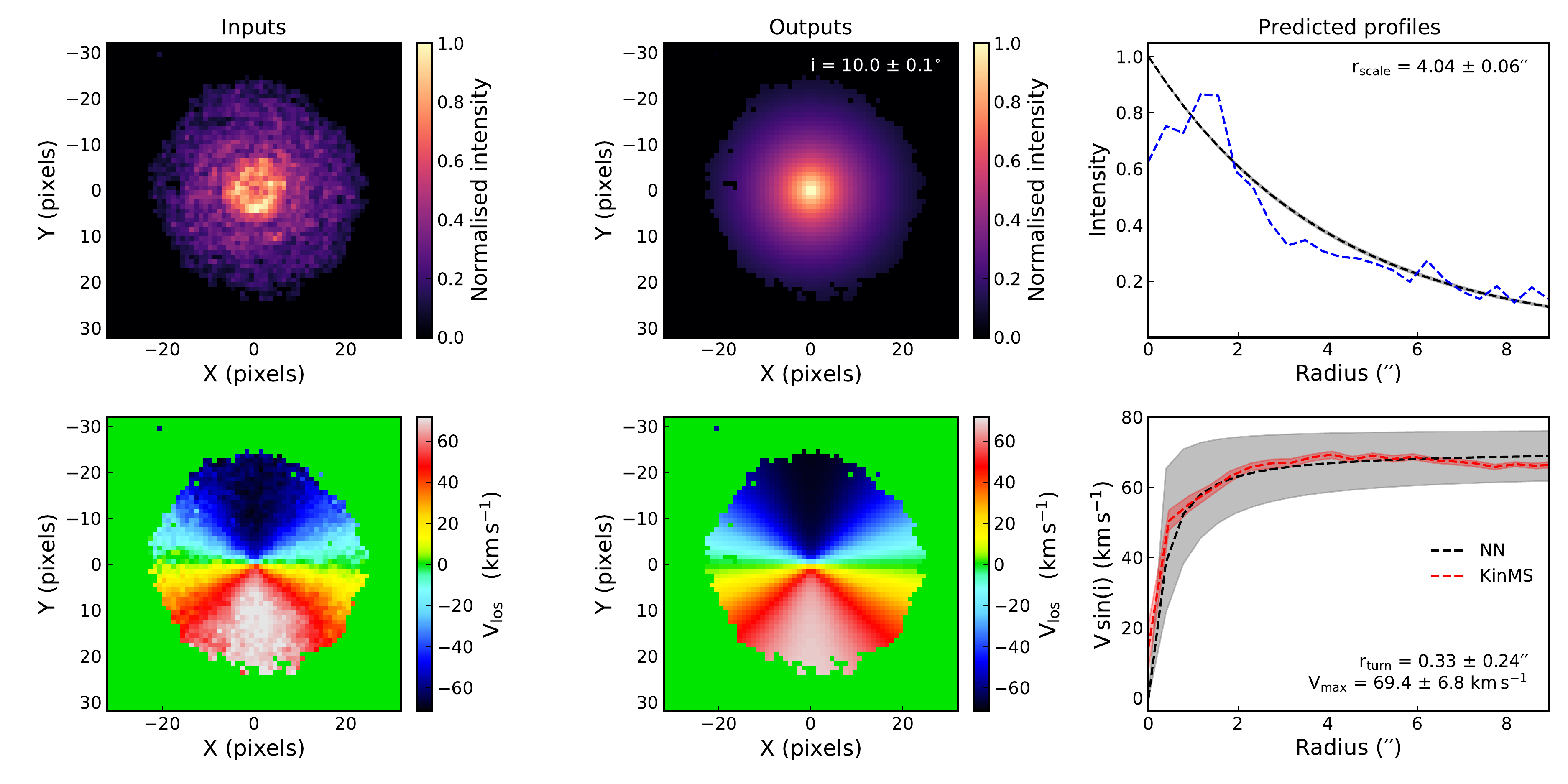}
\caption{An example WISDOM galaxy, NGC 1387, observed in CO and \textcolor{black}{evaluated using} the network in \textit{combined} testing mode. \textcolor{black}{The left and middle columns share $x$ and $y$ axis sizes of 64$\times$64 pixels. In this way we are directly observing the input and output maps of the model. The right column has undergone an $x$-axis rescaling to match observational scales found in the literature. The black dashed lines and grey areas show the mean and standard deviation respectively for profiles predicted by the neural network model.} The blue dashed line shows a major axis cut of the input intensity map. The red dashed line shows the \texttt{KinMS} reconstructed rotation curve. \textcolor{black}{The network predicted parameters are shown as text in the upper-middle, upper-right, and lower-right subplots.} We easily see that this galaxy has an \textcolor{black}{intensity} profile and velocity profile which can be roughly approximated by an exponential and an arctan function respectively, meaning the kinematic parameters are well recovered by the model.}
\label{fig:NGC1387}
\end{figure*}

\subsection{H\textsc{i} examples}\label{sec:VLA}

The primary goal of developing a network like that presented in this paper is to demonstrate the applicability of machine learning to SKA science. As such, in this section we show the network performs well with H\textsc{i} observational data. In order to do this we present \textcolor{black}{two example test galaxies, NGC 2403 and NGC 3198,} observed using the \textbf{V}ery \textbf{L}arge \textbf{A}rray (VLA) as part of \textbf{T}he \textbf{H\textsc{i}} \textbf{N}earby \textbf{G}alaxy \textbf{S}urvey (THINGS) \citep{walter2008}, and showing a diversity of rotation curve shapes. These galaxies are two of 17 THINGS galaxies used for \textit{mixed} training and testing using the network and chosen heuristically for the appearance of their well defined rotating H\textsc{i} disks. The names and publications for the galaxies used in this sample are shown in Table \ref{table:THINGS_info}.

\textcolor{black}{Figure \ref{fig:NGC2403} shows the derived intensity profile and rotation curve for NGC 2403. We include the rotation curve modelled using \texttt{BBarolo} \citep{3DBAROLO} on the datacube (Di Teodoro \& Lelli, private communication). In comparison, we see that the neural network's predicted rotation curve matches closely and so we are convinced that the network is able to recover physical information well. Although the galaxy's intensity profile does not strictly exhibit an exponential form, this has little impact in the recovery of the rotation curve which is the networks primary objective.}

\textcolor{black}{Figure \ref{fig:NGC3198} shows the derived intensity profile and rotation curve for NGC 3198. This galaxy exhibits a \textcolor{black}{mild warp and a flat rotation curve \citep{Gentile} with a slight rise at $\sim200^{\prime\prime}$}. Warped H\textsc{i} discs are not uncommon in the outer regions of galaxies. At present our network architecture is not set up to model these (however one could easily extend the model in order to do so). Again, we include the rotation curve modelled using \texttt{BBarolo} on the datacube (Di Teodoro \& Lelli, private communication) in Figure \ref{fig:NGC3198}.  Crucially, although this warping behaviour is not included in our model, in this case the network still returns reasonable parameter estimations, showing that it could still be usable for parameter estimations across a broadly diverse population of galaxies.}


\subsection{CO examples}\label{sec:WISDOM}

In order to demonstrate the flexibility of this network architecture, we trained a model to recover the kinematic properties of galaxies observed in the CO line using the \textbf{A}tacama \textbf{L}arge \textbf{M}illimeter/submillimeter \textbf{A}rray (ALMA). Our samples are drawn from the mm-\textbf{W}ave \textbf{I}nterferometric \textbf{S}urvey of \textbf{D}ark \textbf{O}bject \textbf{M}asses (WISDOM) project (see Table \ref{table:WISDOM_info} for more information) and have high spatial resolution. Due to the nature of these objects being targeted for their evidence of black hole influence on the gas kinematics, we expect to see small values of a$_\text{V}$ for the sample. As seen in Figure \ref{fig:NGC1387}, this effect is clearly visible, highlighting the predictable behavioural nature of the network. It is also clear in Figure \ref{fig:NGC1387}, that NGC 1387 (FCC184, \citealt{alfocs}, Boyce et al., in prep), an exemplar galaxy from the WISDOM sample, exhibits an exponential \textcolor{black}{intensity} profile which the network can easily recover. 

Such an example demonstrates the \textit{transferable} nature of this network architecture and training style but without the difficulties often associated with traditional transfer learning tasks. This means that such architectures and training styles can be applied to a multitude of different datasets with the possibility of architectural modifications suiting other types of data outside of interferometry and even astronomy.

\subsection{Testing speed}\label{sec:speed}

\textcolor{black}{The network can retrieve a mean field approximation for all learnable parameters, of a single galaxy observation, in 0.0025 seconds on a single Intel(R) Core(TM) i7-6700 CPU core. This time scales linearly with the number of MC dropout samples one wishes to collect (i.e. for a set of 1000 MC dropout samples, a typical test on an individual galaxy would take 2.5 seconds) to generate pseudo-probabilistic distributions. However as the batch throughput size is limited only by the available device memory, it is possible to retrieve values for learnable parameters, and hence MC dropout samples, in the same time frames as listed above for multiple observations. This means that one could potentially return hundreds to thousands of parameterisations and associated pseudo-errors in a matter of seconds.}

\subsection{Caveats}\label{sec:caveats}

There are a few caveats pertaining to the use of the model described in this work. These caveats may impact the way in which users handle the network and the confidence levels associated with parameter estimations.

\textcolor{black}{A key factor in recovering sensible parameterisations using the network is the choice of decoder functions (see \S\ref{subsec:decoder}). In this work we have used simple, general, functions in the form of an exponential (see Equation \ref{eq:SBProf}) and an arctan (see Equation \ref{eq:LOSVel}). However, should one wish to model specific emission line components of galaxies, it would be prudent to adopt more tailored functional forms. For example, it has been shown that H\textsc{i} discs can display depressions in their intensities in their central regions, typically filled by molecular gas \citep{Wong}, for which a truncated Gaussian intensity profile \citep{Martinsson} would be more appropriate when reconstructing the intensity maps. Additionally, when modelling the very outer regions of H\textsc{i} discs, one might consider adopting a more complex multi-parameter function capable of encoding the \textit{sharpness} of the turnover at r$_{\text{turn}}$ and the behaviour of the curve after this point (e.g. \citealt{Rix}), or even declining velocities in the central regions \citep{Federico}. \textcolor{black}{A declining rotation curve would be challenging for the current model to fit (and impossible to fully retrieve)}. However, due to the nature of the loss function chosen in this work (see Equation \ref{eq:MSELoss}), the network will prioritise fitting to the higher velocity regions of galaxies.}  

As described in \S\ref{sec:resolution}, the resolution of input images impacts the ability of the network to correctly predict a$_\text{scale}$, particularly in the high inclination regime. This places constraints on the user's confidence in parameter estimations when working in both the large-beam and high inclination cases combined. Additionally, we can see in Figure \ref{fig:synthetic_cut} that the network struggles to accurately recover inclinations at the very low inclination range. This is a predictable effect caused by the loss of line of sight velocity information for face on disks but again, in the case of survey pipelines, these low inclined galaxies will require additional flagging. In both the aforementioned caveat cases it is worth noting that traditional kinematic modelling methods also struggle to accurately estimate parameters, in particular when working with moment maps. \textcolor{black}{Extensions} of the \textcolor{black}{network's} framework presented here to kinematically model datacubes may alleviate these issues and will be explored in future work. 

\section{Conclusions}\label{sec:conclusion}

We have demonstrated the performance of a neural network model architecture which can be used to recover rotation curves of galaxies from their \textcolor{black}{kinematics}. The model was tested on synthetically generated galaxies as well as observations using both H\textsc{i} and CO emission lines.\\

Testing on synthetically generated galaxies has highlighted the powerful performance of the network as well as areas where the network's performance is sub optimal. For the latter areas we have discussed solutions including: an additional convolution with the restoring beam to counteract the effects of "beam smearing", and flagging high inclination data in a large beam and high inclination regime.\\

Testing observational H\textsc{i} data from THINGS has shown that this style of network is well suited to work with data like that expected from the SKA in the near future. \textcolor{black}{We have shown that the network is capable of estimating velocity curves for discs exhibiting a variety of profiles. In order to do this, we have directly compared the rotation curves estimated by the network to those modelled directly from the cubes using kinematic modelling tools. The network is able to perform adequate recovery of parameters even in cases where it would not be possible to reproduce the true rotation curves. These promising results give us confidence that adopting more flexible decoder functions will extend the applicability of the model for more specific use cases should one wish to model H\textsc{i} discs exclusively.} 

Testing observational CO data from the WISDOM project has shown that the network is suitable \textcolor{black}{for a range of emission line observations}. Unlike traditional ML models, the network architecture and training styles outlined in this work prevent the need for \textit{transfer learning} which is often time consuming and fraught with ungainly challenges associated with systematic properties of training sets. We have shown that the model outlined in this work can recover rotation curves which heuristically match rotation curves extracted from ALMA observations using more time-consuming approaches.\\

As previously stated, improvements to the model architecture in this work include but are not limited to: adapting the model to use more complex \textcolor{black}{intensity} and velocity profiles in the decoder subnet, automatically accounting for large beam effects such as beam smearing and information loss either via systematic offsets in model predictions or via the incorporation of an extra convolutional layer in the decoder subnet, and reintroducing a position angle estimation step. An idealised improvement on the model would be to work directly with interferometric datacubes themselves, \textcolor{black}{or even visibilities}, without the need to generate moment maps prior to training and testing. However, we have found that the discretised nature of channels in interferometric datacubes presents a non-gradient-trackable step in the decoder's reconstruction of datacube inputs. This discontinuity in the gradient tree prevents back propagation via gradient descent and consequently halts model training. \textcolor{black}{We propose adapting this self-supervised approach to work with datacubes as a lucrative avenue of research for challenging current kinematic modelling tools in preparation for the SKA and other upcoming large facilities.}


\section*{Acknowledgements}

This paper has received funding by the Science and Technology Facilities Council as part of the Cardiff, Swansea \& Bristol Centre for Doctoral Training.

We gratefully acknowledge the support of NVIDIA Corporation with the donation of a Titan Xp GPU used for this research. We also thank the anonymous reviewer whose comments and suggestions helped improve this manuscript.

JMD wishes to gratefully acknowledge the help of Dr Federico Lelli for providing rotation curves for THINGS sample galaxies and valuable insight which have both contributed towards improving this paper. 

TAD acknowledges support from the UK Science and Technology Facilities Council through grant ST/S00033X/1.

This research made use of Astropy\footnote{http://www.astropy.org}, a community-developed Python package for Astronomy \citep{astropy_collaboration_astropy:_2013,astropy_collaboration_astropy_2018}, NumPY\footnote{https://numpy.org/} an open source numerical computation library \citep{numpy}, and pandas\footnote{https://pandas.pydata.org/} a data manipulation software library \citep{pandas}.

This paper makes use of data obtained using the Jansky Very Large Array, a component of the National Radio Astronomy Observatory (NRAO). The NRAO  is  a  facility  of the  National  Science  Foundation  operated  under  cooperative agreement by Associated Universities, Inc.

This paper makes use of ALMA data. ALMA is a partnership of the ESO (representing its member states), NSF (USA), and NINS (Japan), together with the NRC (Canada), NSC, ASIAA (Taiwan), and KASI (Republic of Korea), in cooperation with the Republic of Chile. The Joint ALMA Observatory is operated by the ESO, AUI/NRAO, and NAOJ.



\bibliographystyle{mnras}
\bibliography{references} 



\section{Data Availability}

The  data  and  scripts  underlying  this  article  are  available via GitHub, at \url{https://github.com/SpaceMeerkat/Corellia}.



\appendix

\section{Extra material}

\begin{figure}
\includegraphics[width=\linewidth]{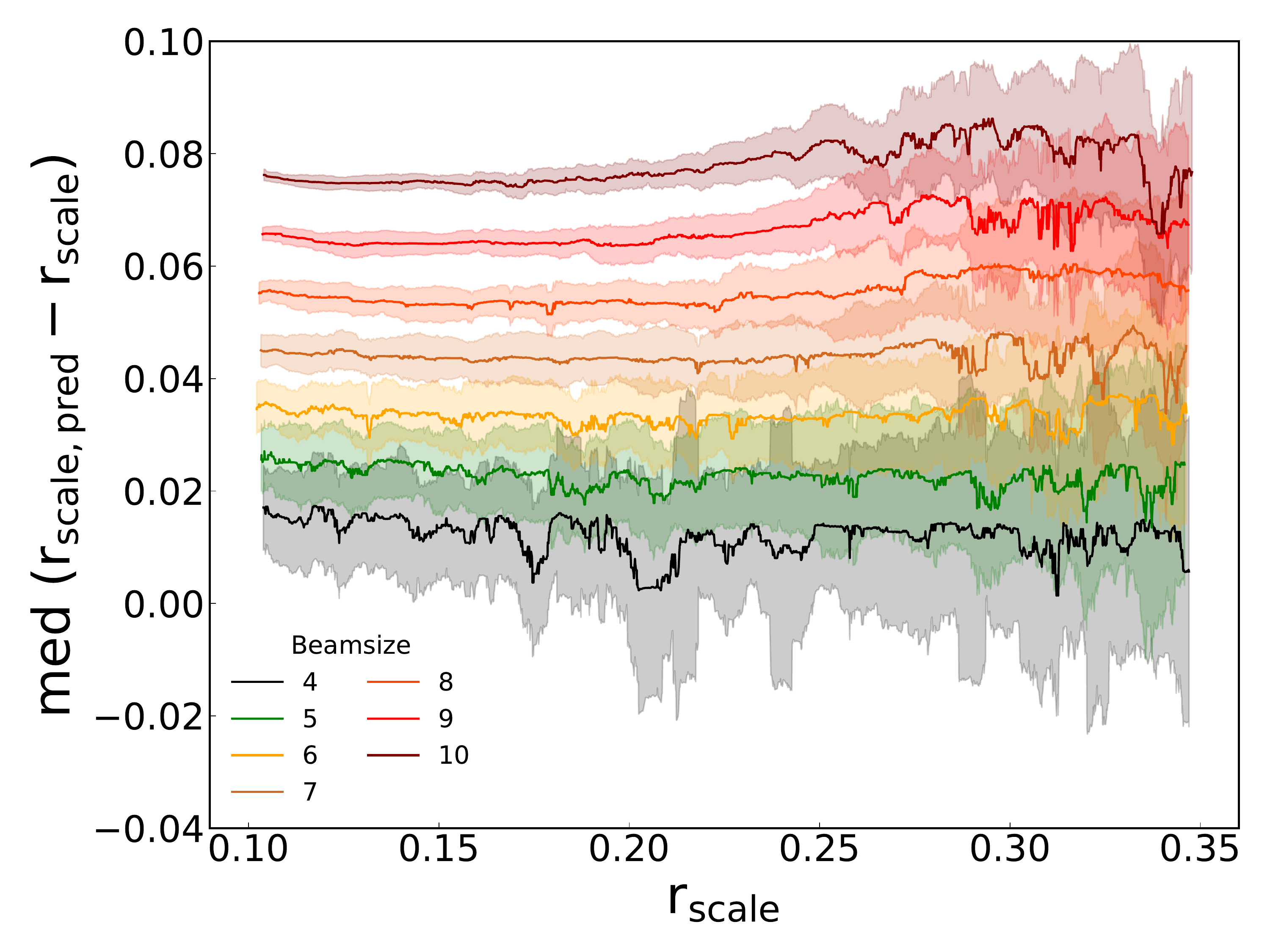}
\caption{The effects of varying the ratio of beam size to galaxy extent. It is clear to see that an increased beam size results in an artificial lengthening of the \textcolor{black}{intensity} profile scale length. It can also be seen that the spread in median offset increases with r$_{\text{scale}}$, which occurs due to information loss as the convolved flux is "smeared" out beyond the field of view. The value of r$_{\text{scale}}$ at which this effect begins to take hold is clearly inversely proportional to the beamsize.}
\label{fig:beam_offsets}
\end{figure}

\begin{table}
\centering
\caption{Information regarding the THINGS sample galaxies used throughout his work. Columns give the following information: \textit{Object}, the target name as given in THINGS project publications, \textit{Publication}, records the relevant publication in which the THINGS targets appear.}
\begin{tabular}{llc}
\hline
\hline
\multicolumn{2}{c}{Object} & Publication \\ \hline
            &            &             \\
DDO 53      & NGC 3621   & \multirow{8}{*}{All from \cite{walter2008,blok}}     \\
NGC 925     & NGC 4736   &            \\
NGC 2403    & NGC 4826   &           \\
NGC 2841    & NGC 5055   &          \\
NGC 2903    & NGC 5236           &          \\
NGC 3184    & NGC 6946   &           \\
NGC 3198    & NGC 7331   &            \\
NGC 3351    & NGC 7793   &            \\
NGC 3521    &    &            \\ \hline
\end{tabular}
\label{table:THINGS_info}
\end{table}

\begin{table}
\centering
\caption{Information regarding the WISDOM project sample used throughout this work. Table columns give the following information: \textit{Object}, the target name as given in WISDOM project publications, \textit{Observation type}, gives the emission line ALMA observed for the target, \textit{Publication}, records the relevant publication in which ALMA observations of the targets appear.}
\begin{tabular}{lll}
\hline
\hline
Object   & Observation type & Publication \\ \hline
         &                                                    & \\
NGC 3665 & $^{12}$CO(2-1)                                     & \cite{NGC3665} \\
NGC 0383 & $^{12}$CO(2-1)                                     & \cite{NGC0383} \\
NGC 0524 & $^{12}$CO(2-1)                                     & \cite{NGC0524} \\
NGC 1387 & $^{12}$CO(2-0)                                     & \cite{alfocs}, Boyce et al. (in prep)  \\
NGC 4429 & $^{12}$CO(3-2)                                     & \cite{NGC4429} \\
NGC 4697 & $^{12}$CO(2-1)                                     & \cite{NGC4697} \\ 
\hline
\end{tabular}
\label{table:WISDOM_info}
\end{table}


\bsp	
\label{lastpage}
\end{document}